# Effects of pulsed and continuous light and heavy ion irradiation on the morphology and electrical properties of Ag+C$_{60}$ and Au+C$_{60}$ composite thin films

G. Ceccio[a*], K. Takahashi[b], Y. Kondo[b], R. Miksova[a], V. Lavrentiev[a], J. Novak[a], E. Stepanovska[a] and J. Vacik[a]

[a]Department of Neutron and Ion Methods, Nuclear Physics Institute of CAS, Rez, Czech Republic;
[b]Department of Electrical Engineering, Nagaoka University of Technology, Nagaoka, Niigata, Japan

*corresponding author, ceccio@ujf.cas.cz

**Abstract**

Metal-organic nanocomposite thin films represent a versatile class of materials whose properties can be effectively tuned through external stimuli. In this study, Ag+C$_{60}$ and Au+C$_{60}$ nanocomposite thin films were briefly investigated to elucidate the effects of ion irradiation on both their morphology and electrical properties. The films were synthesized by co-deposition of noble metals and fullerenes, using ion beam sputtering of metal targets combined with simultaneous thermal evaporation of C$_{60}$. The as-deposited films were characterized by ion beam analysis to determine their composition and element depth distributions. Subsequently, the samples were irradiated at room temperature with either a continuous Ar$^+$ ion beam or a pulsed C$^+$ ion beam, both at an energy of 20 keV and a fluence of $1 \times 10^{15}$ ions/cm$^2$. Irradiation-induced morphological changes were examined by scanning electron microscopy. While the C$^+$-irradiated films retained compact and homogeneous surface morphologies, Ar$^+$ irradiation induced pronounced surface restructuring, resulting in highly corrugated and porous-like surfaces. In addition to morphology, the electrical resistance of the films was measured. The results indicate that C$^+$-irradiated samples exhibit only minor changes in resistivity, whereas Ar$^+$ irradiation strongly affects the electrical properties, with the most significant impact observed for the Au+C$_{60}$ system. The observed changes in electrical resistance closely correlate with the irradiation-induced surface morphology. The measurement results are briefly discussed below.

## Introduction

Fine control over nanomaterials remains a crucial challenge in modern materials science. Considerable attention has been devoted to understanding the dependence of the physical properties of nanostructures on their morphology [1-3]. Even minor variations in structure, morphology, or composition at the nanoscale can result in pronounced macroscopic changes. With advances in irradiation technologies, ion irradiation has emerged as a powerful tool for tailoring nanomaterials, owing to its ability to induce well-defined and reproducible modifications without compromising the macroscopic integrity of the material [4, 5]. Ion beams have been extensively employed for the modification of semiconductors and oxides, as well as for latent track formation in polymers and diamonds [6-9]. The range of applications of ion-modified materials strongly depends on beam parameters such as ion energy, fluence, and mass.

Beyond conventional continuous ion beam irradiation, pulsed ion beams have attracted increasing interest in recent years. Pulsed irradiation delivers ions in short, high-current pulses, creating transient non-equilibrium conditions characterized by rapid energy deposition. These conditions can give rise to modification pathways that differ fundamentally from those observed under continuous irradiation, including enhanced defect clustering, rapid thermal spike formation, and nonlinear response effects. A systematic comparison between continuous and pulsed ion beam irradiation therefore provides valuable insight into radiation-induced modification mechanisms and their impact on material properties.

Metal-organic nanocomposite thin films represent an ideal platform for such investigations due to their structural complexity and high sensitivity to external stimuli [10, 11]. Among organic molecular systems, fullerene $C_{60}$ is particularly attractive because of its unique spherical structure, high electron affinity, excellent thermal stability, and remarkable mechanical robustness [12-15]. Incorporation of noble metals such as Ag and Au into a $C_{60}$ matrix results in hybrid nanocomposite films exhibiting synergistic properties derived from the high electrical conductivity and chemical stability of the metals, as well as metal-organic molecular interactions.

In this work, we investigate the effects of continuous and pulsed ion beam irradiation on co-deposited $Ag+C_{60}$ and $Au+C_{60}$ nanocomposite thin films. By analyzing irradiation-induced changes in surface morphology and electrical properties, we aim to elucidate the role of ion beam parameters in tuning the properties of these hybrid materials. The present study demonstrates the effectiveness of ion beam irradiation as a controlled approach for tailoring nanomaterial properties and establishes metal-organic composite films as promising candidate systems for irradiation-based materials engineering.

**Methodology**

The nanocomposite thin films were synthesized by co-deposition using the simultaneous application of ion beam sputtering (IBS) of metal targets and thermal evaporation of $C_{60}$ powder onto Si(100) substrates (TedPella). Film deposition was carried out at the Low Energy Ion Facility (LEIF), a laboratory-assembled multipurpose system dedicated to the synthesis and modification of materials using keV ion beams, operating at a base pressure of $2.5 \times 10^{-6}$ mbar [16, 17]. The facility provides intense gas ion beams generated by a duoplasmatron ion source, delivering beam currents up to the mA range and energies between 100 eV and 35 keV.

For the present experiments, an $Ar^+$ ion beam with an extraction voltage of 20 kV and a beam current of 0.5 mA (measured at the target position) was employed. The beam was focused onto high-purity noble metal targets mounted on a rotating holder and inclined at 45° with respect to the $Ar^+$ beam direction. Gold targets (50 mm × 1 mm, 4N purity, Safina) were used for the first set of samples $Au+C_{60}$, while silver targets (50 mm × 3 mm, 4N purity, MSE Supplies) were used for the second set $Ag+C_{60}$.

For co-deposition, a laboratory-built thermal evaporator equipped with a quartz crucible was positioned directly beneath the substrate holder and filled with $C_{60}$ powder (3N purity, Nanografi). The evaporator temperature was remotely controlled using a digital controller and maintained at 450 °C, as measured by a type K thermocouple. Throughout the deposition process, the Si substrates were kept at room temperature.

After deposition, the films were analyzed by ion beam analysis (IBA) techniques to determine heavy element distributions, film thickness, and to verify the absence of contaminants. The IBA measurements were performed using the Tandetron MC 4130 accelerator at the CANAM infrastructure of the Nuclear Physics Institute [18]. Rutherford backscattering spectrometry (RBS) was employed for the analysis [19].

He$^+$ ions with an energy of 2 MeV were used for the identification of heavier elements, while H$^+$ ions at 1.735 MeV were selected to enhance sensitivity to carbon. The samples were measured at an incidence angle of 7° to minimize channeling effects. Backscattered ions were detected using an ORTEC ULTRA-series detector with a 50 mm$^2$ active area and a 300 µm depletion layer, positioned at a scattering angle of 170° out of the plane in Cornell geometry. Data evaluation was carried out using the SIMNRA simulation code [20].

Continuous ion beam irradiation of the hybrid films was performed using the LEIF system, with samples mounted in the target holder orthogonally to the Ar$^+$ beam at the position of the sputtering target. Irradiation was conducted using 20 keV Ar$^+$ ions at a reduced beam current density of 100 µA/cm$^2$. Selected samples were irradiated to a total fluence of $1 \times 10^{15}$ Ar$^+$ ions/cm$^2$.

Pulsed ion irradiation was carried out using a carbon ion beam generated by a laser ion source [21], applying the same total ion fluence. Carbon ions were produced by laser ablation of a graphite target using a Q-switched Nd laser operating at a wavelength of 532 nm, with a pulse duration of 17 ns and a pulse energy of 223 mJ. The laser beam was focused onto the target using a lens with a focal length of 800 mm, with an incidence angle of 29° relative to the target normal. The laser spot diameter was approximately 4 mm, corresponding to an estimated laser intensity of $1 \times 10^8$ W/cm$^2$. Carbon ions were extracted through a 20 mm diameter aperture positioned 100 mm from the target surface and subsequently accelerated to an energy of 20 keV. The resulting pulsed ion beam exhibited a peak current density of 700 µA/cm$^2$ and a full width at half maximum (FWHM) pulse duration of 6 µs. The estimated ion fluence per pulse was $3 \times 10^{10}$ ions/cm$^2$. The laser was operated at a repetition rate of 10 Hz, yielding a total carbon ion fluence of approximately $1 \times 10^{15}$ ions/cm$^2$ after one hour of irradiation.

The surface morphology of the films, both as-deposited and after ion irradiation, was examined using scanning electron microscopy (SEM, Hitachi SU8230) and Atomic Force Macroscopy (employing NT-MDT NTEGRA AURA scanning probe microscope). Electrical resistance measurements were performed to evaluate changes induced by irradiation using a Keithley 2182A nanovoltmeter in conjunction with a Keithley 6221 DC current source operating in a galvanostatic configuration. Measurements were conducted under ambient conditions using the standard two-point probe technique.

**Results**

The Ion Beam Analysis (IBA) techniques employed to analyze the prepared hybrid samples Ag+C$_{60}$/Si and Au+C$_{60}$/Si enable the determination of elemental composition and depth profiles, as well as the detection of possible contaminants or oxidation within the hybrid layers. The results of the RBS analysis, including experimental spectra and SIMNRA fitting, are presented in Fig. 1 for the Ag+C$_{60}$ (a, c) and Au+C$_{60}$ (b, d) films. The figure displays energy spectra acquired using 2 MeV alpha particles and 1.735 MeV proton beams. The combined RBS analysis allows determination of the depth distribution of the constituent elements. As can be seen, the spectra reveal the presence of only two elements: the noble metal (Ag or Au) and carbon. Their contributions are identified by the corresponding peaks: channels 675 and 725 correspond to Ag and Au, respectively (panels a and b), while channel 510 corresponds to carbon (panels c and d).

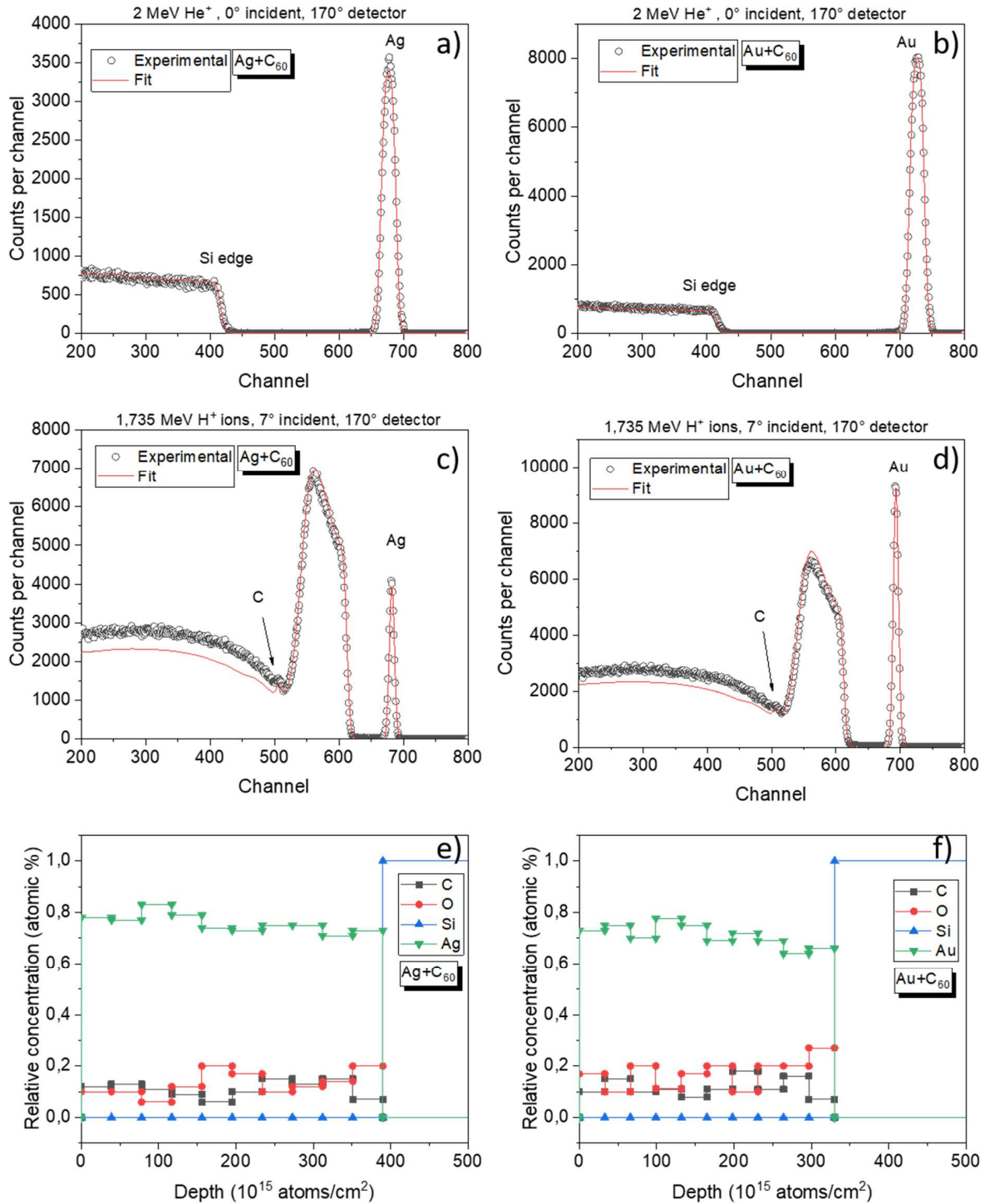

*Figure 1. Panels a) and b) show RBS results on Ag+C$_{60}$ and Au+C$_{60}$ samples obtained with a 2MeV He beam. Panels c) and d) show the RBS analysis using a 1.735 MeV H beam. Lower panels e) and f) show the relative concentration of elements vs depth.*

In the Ag+C$_{60}$ film, the constituent elements exhibit a homogeneous depth distribution, with the Ag content ranging between 71 and 83 at.%. The carbon content is also uniform throughout the analyzed depth and amounts to approximately 10–15 at.%. In the Au+$C_{60}$ film, the Au concentration fluctuates around lower values, between 64 and 78 at. %, while the carbon content ranges from 8 to 20 at.%. Oxygen is not explicitly visible in the RBS spectra, as its signal overlaps with and lies below that of the Si substrate. However, the oxygen content can be estimated through precise spectral simulations based on atomic percentages. Since the C and Ag (or Au) concentrations are well defined, the remaining fraction in the atomic composition can be attributed to oxygen. The constant depth profile of oxygen suggests that it does not originate from surface oxidation, but rather from unintentional contamination during the deposition process.

The characterization of the pristine surfaces of the hybrid films is presented in Fig. 2, which shows SEM and AFM micrographs of the as-deposited Ag+$C_{60}$ (left) and Au+$C_{60}$ (right) samples. As shown in Figs. 2, both films exhibit flat and homogeneous surfaces, with no discernible surface features or nanostructure formation. The average surface roughness measured using AFM result to be 3.732 nm for the Ag+$C_{60}$ and 2.496 nm for the Au+$C_{60}$.

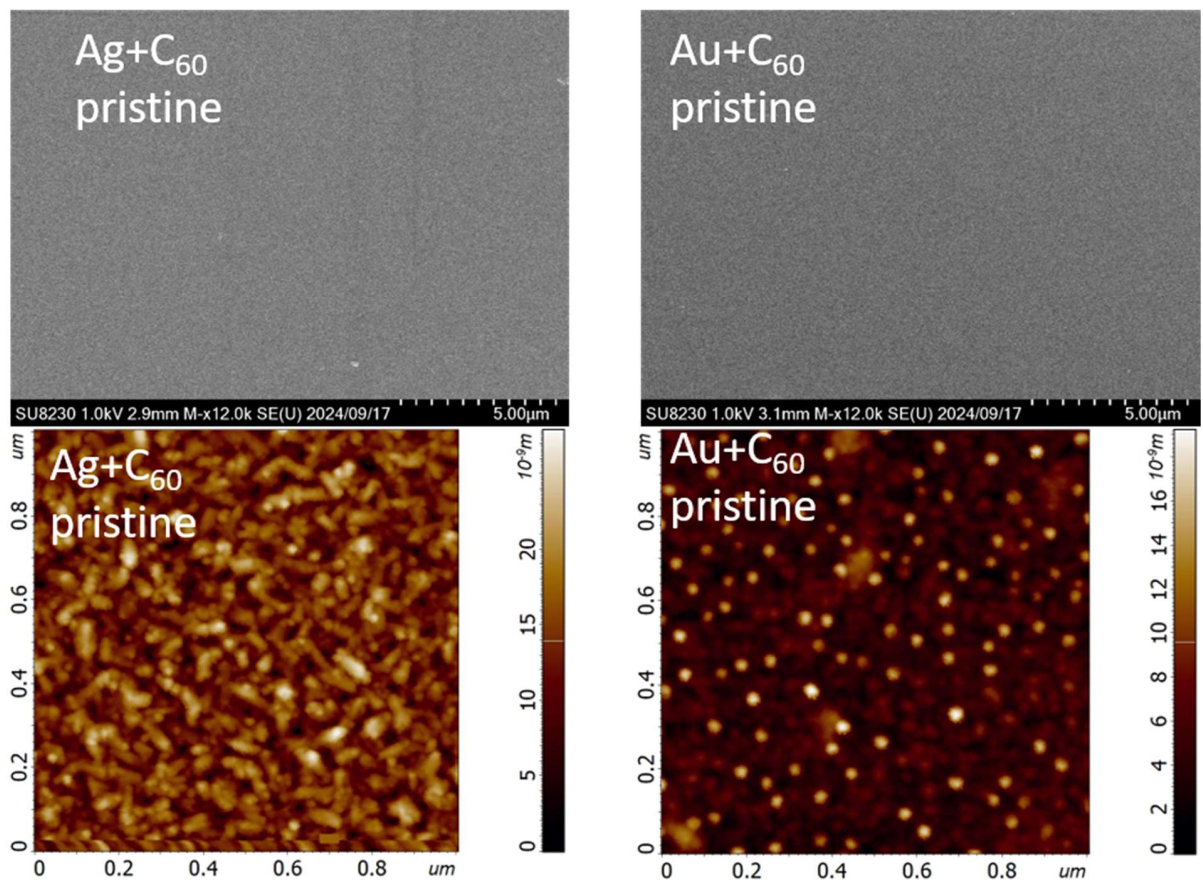

*Figure 2. SEM micrographs of pristine samples Ag+$C_{60}$ (left) and Au+$C_{60}$ (right) in as-deposited state.*

As indicated above, the samples were irradiated at the same fluence of $1 \times 10^{15}$ ions/cm$^2$ using a continuous Ar$^+$ beam and a pulsed C$^+$ beam, both with an energy of 20 keV. SRIM simulations were employed to estimate the ion penetration depth and depth distribution within the films, as well as to assess differences in the damage profiles induced by light and heavy ions. It should be noted that the SRIM simulations do not account for the specific characteristics of the pulsed C$^+$ beam. Fig. 3 illustrates the differences between the simulated ion depth distributions.

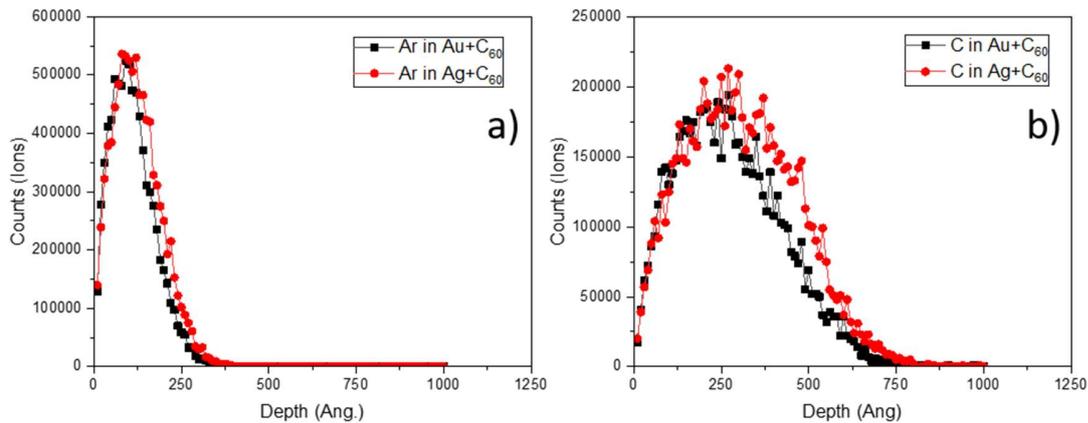

*Figure 3. SRIM simulated distributions of 20 keV $Ar^+$ (a) and $C^+$ (b) ions in the $Ag+C_{60}$ and $Au+C_{60}$ hybrid films.*

A clear difference between the two simulated profiles is observed. $Ar^+$ ions exhibit a narrower depth distribution with a smaller projected range and reduced straggling, indicating more localized energy deposition and damage generation. In contrast, $C^+$ ions penetrate deeper into the hybrid layers and display a broader depth distribution, resulting in a more homogeneous energy deposition over a larger depth. These differences arise from the distinct energy transfer mechanisms of the two ion species. Owing to its higher mass, Ar primarily loses energy through nuclear stopping, leading to dense collision cascades and a higher defect density near its projected range. Conversely, the lighter C ions undergo relatively stronger electronic stopping, which enables deeper penetration and a more spatially distributed modification of the film structure.

Fig. 4 shows micrographs of the $Ag+C_{60}$ and $Au+C_{60}$ films irradiated with a pulsed $C^+$ ion beam. Both films exhibit compact and continuous surfaces, with no evidence of cracks, delamination, or other macroscopic damage. The surfaces display a fine-grained nanostructure, with contrast variations homogeneously distributed across the films. After irradiation the roughness reduced to 1.487 nm for $Ag+C_{60}$ and to 1.392 nm for $Au+C_{60}$ showing a very soft sputtering effect at atomic level. The absence of visible surface damage or localized features in either sample suggests that $C^+$ ion irradiation at 20 keV and a fluence of $1 \times 10^{15}$ ions/cm$^2$ does not induce severe surface damage or large-scale restructuring. This observation is consistent with the larger ion penetration depth predicted by the SRIM simulations shown in Fig. 3b.

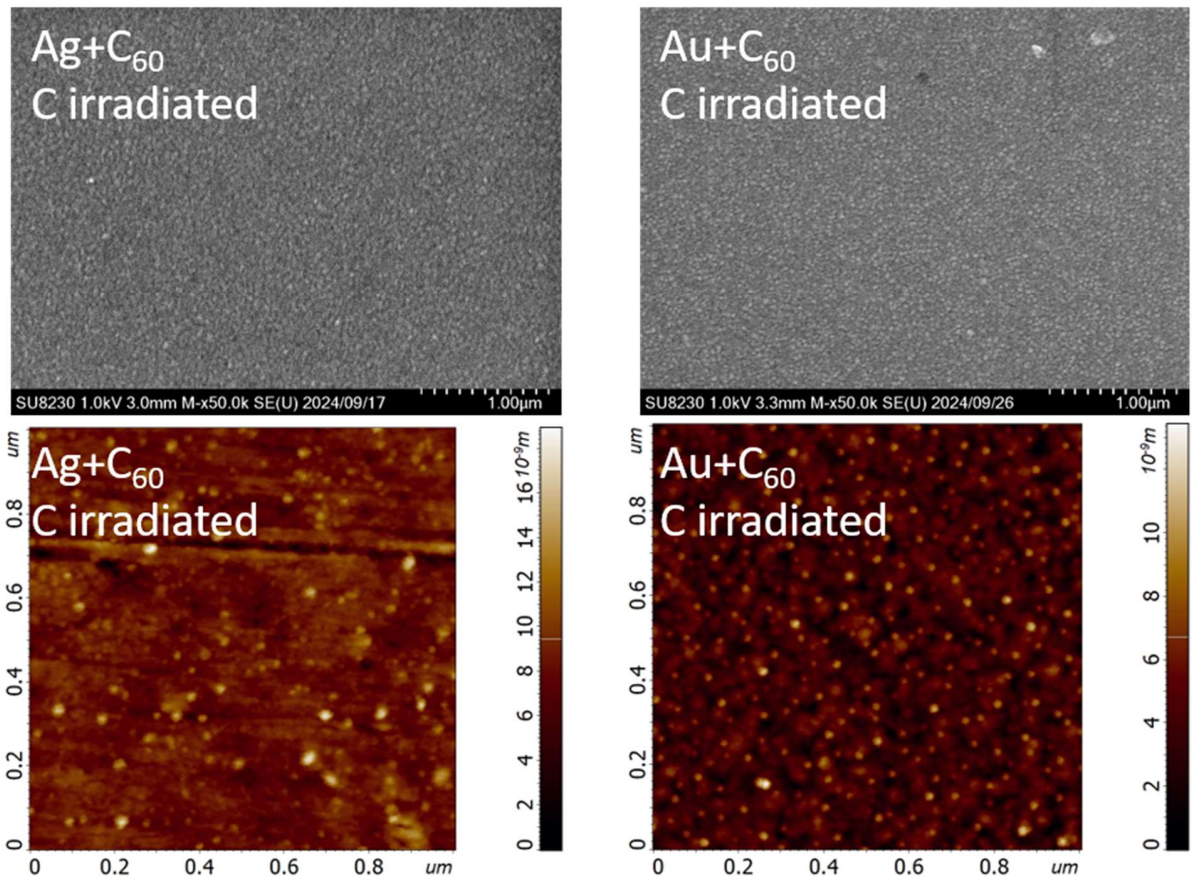

*Figure 4. SEM images of C-irradiated Ag+C$_{60}$ and Au+C$_{60}$ films, acquired at 50k× magnification.*

In contrast to the C-irradiated films, the Ar-irradiated samples exhibit pronounced surface morphological restructuring. As shown in Fig. 5, the SEM images reveal a highly corrugated surface with a porous-like morphology, characterized by a complex network of nanoparticles. These features indicate a substantial modification of the near-surface region induced by Ar$^+$ ion irradiation. AFM micrograph confirmed the increase of roughness with higher value of 9.180 nm for Ag+C$_{60}$ and 9.576 nm for Au+C$_{60}$. Is possible to observe also a larger separation and voids in the Au+C$_{60}$ sample, showing a strong removal of material.

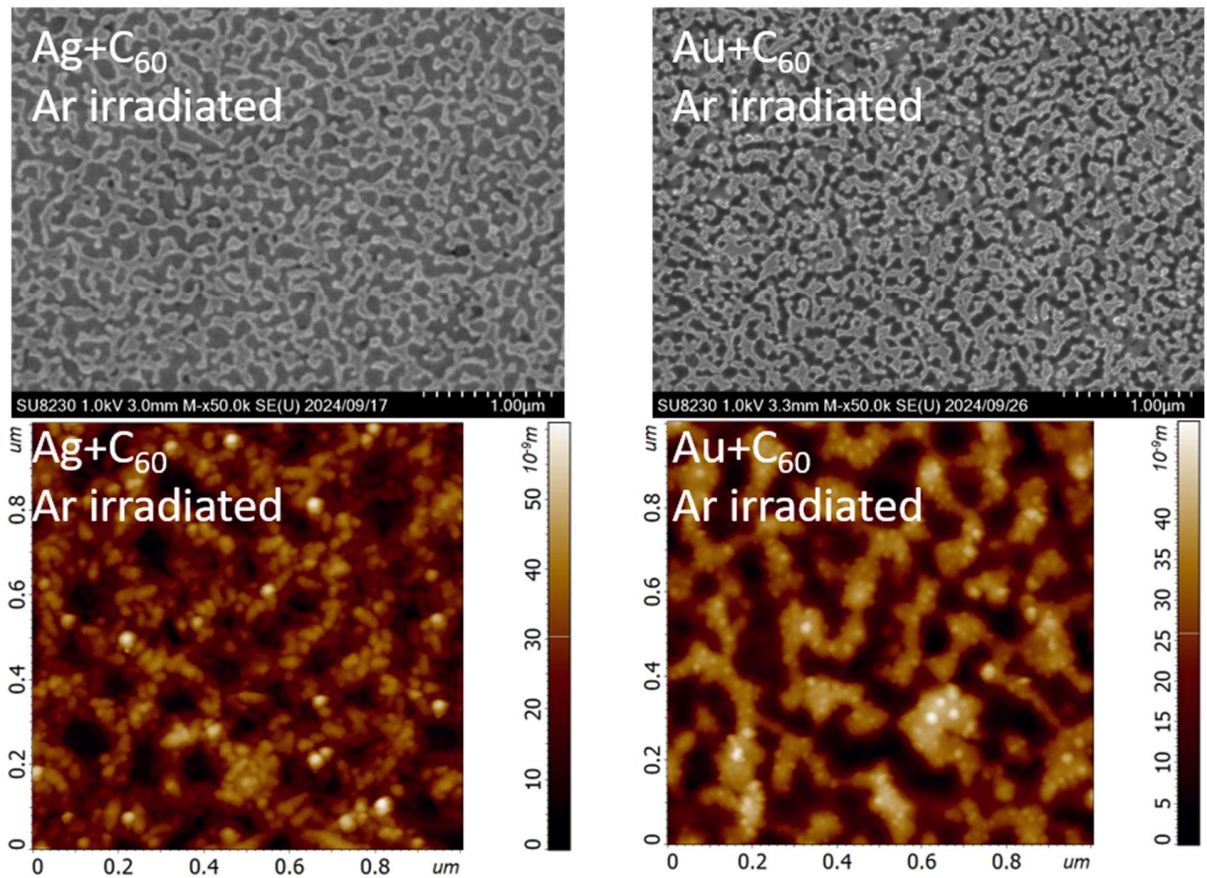

*Figure 5. SEM images of Ar- or C-irradiated Ag+C$_{60}$ and Au+C$_{60}$ films, acquired at 50k× magnification.*

The Ar-irradiated Ag+C$_{60}$ layer exhibits a surface structure consisting of a network of irregular, interconnected features, indicative of pronounced surface roughening and material redistribution. Similarly, the Ar-irradiated Au+C$_{60}$ sample displays a porous morphology composed of irregular, albeit somewhat more compact, nanoparticle formations. The substantial transformation of the surface in both samples can be attributed to the intense collision cascades generated by Ar$^+$ ions interacting with the constituent atoms of the layer during irradiation. This process leads to significant sputtering and the formation of nanodomains with complex shapes anchored to the substrate. In the case of Ag+C$_{60}$, most nanodomains are interconnected, whereas in Au+C$_{60}$ they are predominantly isolated. A clear distinction is therefore observed between the effects of Ar$^+$ and C$^+$ ion irradiation: while C$^+$ irradiation preserves a relatively smooth and compact surface morphology, Ar$^+$ irradiation induces strong near-surface restructuring and pronounced roughening.

RBS measurements were performed after the irradiations, to investigate the changes in composition and reduction in material amount. Fig. 6 a) and b) show the RBS spectra obtained for Ag+C$_{60}$ and Au+C$_{60}$ respectively.

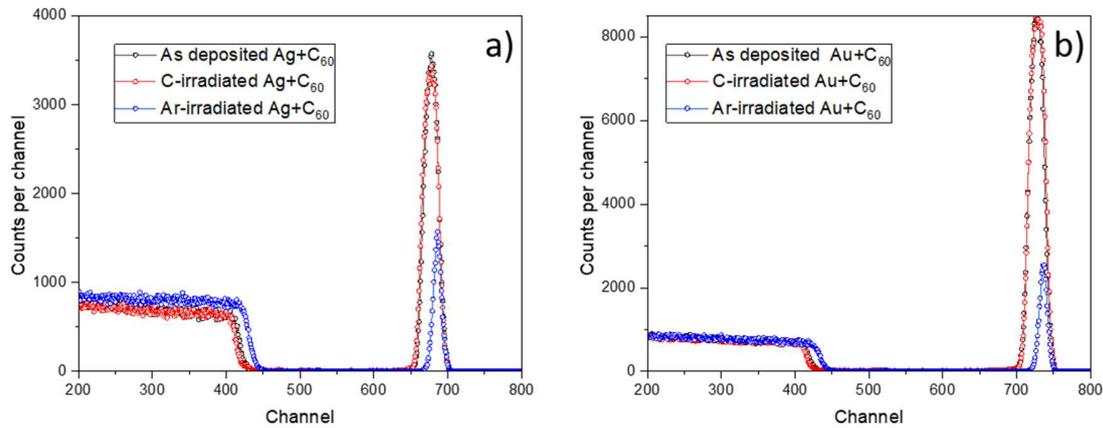

*Figure 1. Panels a) and b) show RBS results on Ag+C$_{60}$ and Au+C$_{60}$ samples obtained after irradiation with pulsed C beam and after continuous Ar beam.*

As showed in the Fig. 7 the films remained almost the same after the C irradiation, as also confirmed by SEM microscopy. After the Ar irradiation, both Ag+C$_{60}$ and Au+C$_{60}$ are subjected to a strong reduction of materials. In particular, from the Ag and Au peaks in the spectra is possible to observe a reduction that can be calculated to be to the 26.38% for the Ag and to the 15.27% for Au. The significant surface restructuring observed in the Ar-irradiated samples is expected to have a significant impact on their electrical properties. The development of a roughened surface increases the surface-to-volume ratio, which can enhance sensitivity to atmospheric gases and alter the surface electrical resistance. Modifications may also occur in the C-irradiated samples, where structural defects in the surface and subsurface regions could influence charge-transfer properties. Fig. 7 presents the results of electrical resistance measurements obtained using the two-point probe method on the sample surfaces. Electrical resistance measurements revealed that all pristine and C-irradiated samples exhibit low resistance values, indicative of metal-dominated conductivity, consistent with the presence of highly conductive Ag or Au components. Similarly, Ar-irradiated Ag+C$_{60}$ films remain conductive, suggesting that metal percolation pathways are largely preserved despite the pronounced surface restructuring observed by SEM. In contrast, the Ar-irradiated Au+C$_{60}$ sample exhibits a significant increase in resistance, likely due to morphological and structural disorder that disrupts conductive pathways. The differing responses of Au- and Ag-based systems underscore the greater sensitivity of the Au system to irradiation-induced modifications, highlighting its potential relevance for gas-sensing applications.

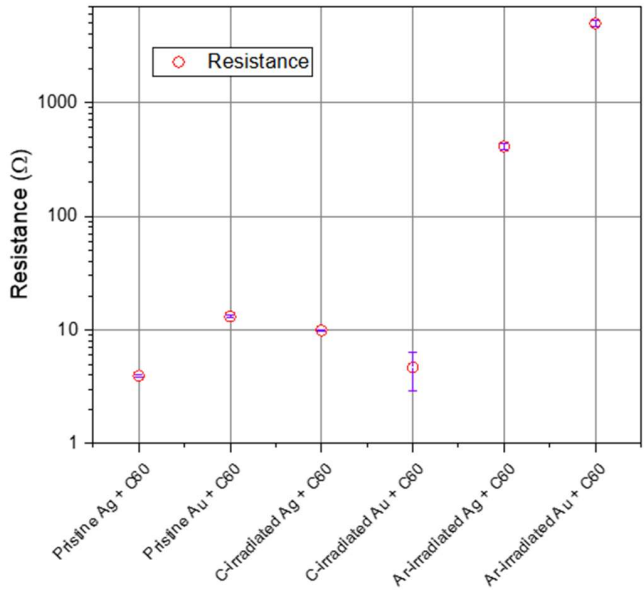

*Figure 7. Electrical resistance measured by 2-point method for pristine, C-irradiated, and Ar-irradiated Ag+$C_{60}$ and Au+$C_{60}$ hybrid samples.*

**Conclusions**

Ag+$C_{60}$ and Au+$C_{60}$ nanocomposite thin films were synthesized by co-deposition of their constituent elements using thermal evaporation (of $C_{60}$) and ion beam sputtering (of Ag and Au, respectively). The prepared samples were subsequently irradiated with ions in both continuous (by $Ar^+$) and pulsed (by $C^+$) regimes. The effects of irradiation were evaluated in terms of surface morphology evolution and changes in electrical resistance. A clear ion-dependent response was observed, consistent with SRIM simulations: heavy $Ar^+$ ions deposit their energy within a narrow near-surface region, whereas lighter $C^+$ ions penetrate deeper into the films, resulting in a more distributed energy deposition. These differences are directly reflected in the observed surface morphologies. $C^+$ irradiation preserves a compact and homogeneous surface structure in both sets of samples, while $Ar^+$ irradiation leads to pronounced roughness and corrugated structures. Electrical resistance measurements indicate that pristine and $C^+$ irradiated films maintain high conductivity, consistent with their metallic character. In contrast, $Ar^+$ irradiation of the Au+$C_{60}$ film causes a significant increase in resistance, suggesting substantial disruption of the conductive layer due to heavy-ion-induced structural disorder.

Overall, this study demonstrates the potential to tune the properties of these nanocomposite systems and highlights the possibility of achieving different effects by selecting ions with distinct characteristics.

**Acknowledgment**

The authors acknowledge the financial support from the MEYS CR, Project OP JAK FerrMion, No. CZ.02.01.01/00/22_008/0004591. The authors acknowledged also the support of Czech Academy of Science Mobility Plus Project, Grant No. JSPS-24-12 and JSPS Bilateral Program Number JPJSBP120242501. Measurements were carried out at the CANAM infrastructure of the NPI CAS Rez under project LM 2015056.